\documentclass[%
 reprint,showpacs,showkeys, 
nofootinbib,
 amsmath,amssymb,
 aps,
]{revtex4-2}

\usepackage{graphicx}
\usepackage{dcolumn}
\usepackage{bm}

\begin{document}
	
\title{Most Probable Energy Distributions of Particles with Hierarchial Structures }

\author{Michael Romanovsky}
\email[]{slon@kapella.gpi.ru}

\affiliation{Science and Innovation Company, 24 Bolshaya Ordynka Street, Moscow, Russia \\ MIREA - Russian Technological University, 78 Vernadsky Avenue, Moscow, Russia}

\date{\today}

\begin{abstract}
Similarly to the derivation of the Gibbs-Boltzmann distribution for structureless indistinguishable particles, we consider multi-particle systems some of which are contained (or delimited) inside others (Problem 1), as well as systems of particles delimited within other particles, which, in turn, are delimited yet inside another kind of particles (Problem 2). Under the natural assumptions concerning the conservation laws, such as the conservation of the total number of particles, total energy, etc, the problem of the most probable energy distributions is studied in a combinatorial formulation, with the obtained distributions treated as externally observable. For Problem 1, the particle distributions over maximal and minimal energies are also established and shown to coincide with the ones found in the framework of the original combinatorial treatment. The results for Problems 1 and 2 can be interpreted as a sorting of lower-level particles based on their statistical mechanics behavior observed in various experiments. An effective Pauli principle arises in a non-contradictory manner in one-particle observations for the complete probabilistic Problem 1 in the combinatorial formulation as well as in the problem of distributions over maximal and minimal energies. Several of the calculated distributions describe particles having no negative energy states. 
\end{abstract}



\maketitle

\section{Introduction}

It is well-known that, under the natural assumptions that the total number of particles is finite and that there exists a mean value of the energy of a particle, a system of indistinguishable identical particles with a discrete energy spectrum sustains, as the most probable, an exponential probability density law given by 
$ W({\epsilon}_i )={(1 / {T_{eff}})} e^{- {{\epsilon}_i} / {T_{eff}}}$ for the particles being in {\it i}-th energy state ${{\epsilon}_i}$. Here $T_{eff}$ is a normalization coefficient standing for a mean energy or an effective temperature of the system ~\cite{isihara1971statistical, landau1980lifshitz}. Let us designate the result as "Problem 0". The thermodynamics characteristics and laws for such systems of structureless particles are well-known. 

Systems of greater complexity, consisting of indistinguishable particles having inner structures, may be subject to modified thermodynamic relations, as in the plasma case ~\cite{romanovsky2006elementary}, and moreover exhibit new effects like negative friction ~\cite{romanczuk2012active} (such particles are routinely described as active). Thereby, the question arises naturally about the distributions of the particles over energies in the most general case, where certain groups of particles are delimited from others or some of the particles, called the higher-level ones in what follows, contain others which are indistinguishable lower-level ones. 

This problem has also been considered (for example, see ~\cite{isihara1971statistical, landau1980lifshitz}. Under the additional {\it a priori} assumption that the numbers of lower-level particles within the higher-level ones vary, the most probable distribution is shaped by a large canonical ensemble. It is not postulated here that the higher-level particles must be identical. Let us designate the above as "Problem 0`". The consideration does not have to be limited to particles and energy distributions a system internally carrying certain structures which affect its external characteristics may be considered as alternative example. The role of the energy distribution for the lower-level particles is played by a distribution of the lower-level system over one of its characteristics. 

It appears natural to switch to a more generalized formulation of the problem, taking into account the observable indistinguishability of the higher-level particles. Designate this formulation as "Problem 1" and note that it implies constructing the most probable distribution of lower-level particles over energies. Multiplicity of such distributions should not be excluded. First, those would be the presumed "genuine" lower-level particle distributions over energies within a single higher-level particle. It should be established at this point whether they are observable and, if they are, what experiments it takes to register them. Secondly, the question is which distributions are observable and what sets them apart from the genuine ones. Problem 0 clearly reduces to a limit of Problem 1 if the higher-level particle (under certain conditions) contains a single lower-level one. 

The formulation acquires additional complexity when the particles have internal structures composed of other particles, which further contain yet another type of species. The above may be appropriately termed "Problem 2". The common pertinent illustration is represented by particles contained in boxes (for example, see ~\cite{isihara1971statistical, landau1980lifshitz}). The boxes being enclosed in other boxes, we arrive at Problem 2. To an extent, atomic nuclei consisting of nucleons, which, in their turn, are built of quarks, can be viewed as some illustration of Problem 2.

The combinatorial problem about the number of combinations in the case where identical objects are distributed over groups is suggestive of a solution to Problem 1. We further demonstrate that considering the distribution of minimal and maximal energies for lower-level particles in a higher-level sample yields the same result. A solution to Problem 2 is spelled out in what follows. 

\section{Problem 1. Combinatorial Formulation }

(see ~\cite{romanovsky2021statistical}). Suppose $m_i$ is a set of higher-level particles such that $m_1$ of these indistinguishable species contain $p_1$ lower-level particles \footnote{Replacing the expression "higher-level particles" with the word "strings", and the expression "lower-level particles" with the word "mushrooms", one arrives at the problem about drying mushrooms ~\cite{vilenkin1969, vilenkin1975} or at the cyclic permutations problem ~\cite{stanley1997enumerative}.},  $m_2$ similarly contain $p_2$, etc. up to the $N$-th lower-level particle. Here $n$ is the overall number of arrangements of the lower-level particles over the higher-level ones. For the total number of lower-level particles $N$, we have 

\begin{equation}
N = {{\sum}_{i=1}^n}{m_i}{p_i} 
 \label{eq_1}
\end{equation}

with a distribution over a total of $M$ higher-level particles

\begin{equation}
M = {{\sum}_{i=1}^n}{m_i} 
 \label{eq_2}
\end{equation}

The total number of permutations of the lower-level particles over all the higher-level particles makes $W_{10}$ given by ~\cite{vilenkin1969}

\begin{equation}
W_{10} = {N! \over{{{\prod}_{i=1}^n}{(p_i !)}^{m_i}}} 
 \label{eq_3a}
\end{equation}

If the higher-level particles being indistinguishable as mentioned above, we have 

\begin{equation}
W_{11} = {N! \over{{{\prod}_{i=1}^n}{(p_i !)}^{m_i}{{\prod}_{i=1}^n}{(m_i !)}}} 
 \label{eq_3b}
\end{equation}

Rather, the expression for $ln W_{11}$ is typically examined 

\begin{equation}
ln W_{11} = ln N! - {{\sum}_{i=1}^n}{m_i}{ln (p_i !)} - {{\sum}_{i=1}^n}{ln (m_i !)} 
 \label{eq_{4a}}
\end{equation}

Note that if $m_i {\equiv} 1$, the formulation of the problem of calculating the most probable distribution, as defined by conditions ~(\ref{eq_1}), ~(\ref{eq_2}),  ~(\ref{eq_{4a}}), becomes analogous to the formulation of Problem 0\footnote{Allowing variations of $N$ in a large canonical ensemble problem, we can reduce it to Problem 1. The process has never been implemented (for example, see ~\cite{stanley1997enumerative}).}. 

Suppose every lower-level particle may be in a state with some energy  ${{\epsilon}_i}$. To find the most probable distribution over these energies for arbitrary $m_i$, one should take into account that, in addition to Eqs. ~(\ref{eq_3b}) or  ~(\ref{eq_{4a}}) and the relations expressed by Eqs. ~(\ref{eq_1}) and  ~(\ref{eq_2}), the condition must be met that the total energy of the system is finite 

\begin{equation}
E = {{\sum}_{i=1}^n}{m_i}{p_i}{{\epsilon}_i} 
 \label{eq_5}
\end{equation}

In practice, this means that there exists a mean "energy" per one lower-level particle $<{\epsilon}>$. 

Let us solve Problem 1. As the initial step, consider the variation of Eq. ~(\ref{eq_{4a}}), incorporating the simplest constraints imposed by Eqs.~(\ref{eq_1}), ~(\ref{eq_2}),  and ~(\ref{eq_5}). To this end, the terms $ {\alpha} N + {\beta}M + {\gamma}E$, with Lagrange multipliers $\alpha$, $\beta$, and $\gamma$ should be added to the right-hand side of ~(\ref{eq_{4a}}):

\begin{eqnarray}
ln W_{1} = ln N! - {{\sum}_{i=1}^n}{m_i}{ln (p_i !)} - {{\sum}_{i=1}^n}{ln (m_i !)} + 
\nonumber \\ + {\alpha}{{\sum}_{i=1}^n}{m_i}{p_i} + {\beta} {{\sum}_{i=1}^n}{m_i} +{\gamma} {{\sum}_{i=1}^n}{m_i}{p_i}{{\epsilon}_i}
 \label{eq_{4b}}
\end{eqnarray}

We also use Stirling's lowest approximation for $m !$ and $p !$, and perform the variation in $ \delta p_i$ and $\delta m_i$. The expression for the variation of the most probable distribution $ln W_1$ becomes

\begin{eqnarray}
{\delta} ln W_{1} = {{\sum}_{i=1}^n} [- ({\delta}{m_i}) p_i (ln p_i -1) -
\nonumber \\
 - ({\delta} p_i) m_i ln p_i - ({\delta} m_i) ln m_i + 
\nonumber \\ + {\beta} ({\delta} m_i) + {\alpha} ({\delta} m_i) p_i + 
\nonumber \\
+ {\alpha} m_i ({\delta} p_i) +{\gamma} ({\delta} m_i) p_i{{\epsilon}_i} + 
\nonumber \\
+ {\gamma} ({\delta} p_i) m_i{{\epsilon}_i}] = 0
 \label{eq_6}
\end{eqnarray}

We further collect the terms related to the independent variations $\delta p_i$ and $\delta m_i$. The first equation of the set is obtained upon the cancellation of the non-zero factors $\delta p_i$ and $m_i$

\begin{equation}
ln p_i = {\alpha} + {\gamma}{{\epsilon}_i} 
 \label{eq_7}
\end{equation}

For the second variation, the cancelation of the non-zero quantity $\delta m_i$ and the use of ~(\ref{eq_7}) show that

\begin{equation}
ln m_i = {\beta} + p_i 
 \label{eq_8}
\end{equation}

Eqs. ~(\ref{eq_7}) and ~(\ref{eq_8}) provide a solution to Problem 1. 

A solution to Eq. ~(\ref{eq_7}) corresponds exactly to the representation embodied in the ordinary Boltzmann-Gibbs distribution (see the Introduction, ~\cite{isihara1971statistical}) and to the solution of Problem 0. The difference from the solution to Problem 0 is that the latter involves no variation in the number of the lower-level particles. Easily, the solutions to Eqs. ~(\ref{eq_7}) and ~(\ref{eq_8}) are

\begin{equation}
p_i = e^{{\alpha} + {\gamma}{{\epsilon}_i}} 
 \label{eq_9}
\end{equation}

and

\begin{equation}
 m_i = e^{{\beta} + p_i} 
 \label{eq_10}
\end{equation}

Thus, the genuine function of the most probable energy distribution of lower-level particles is the same as in the most probable distribution in Problem 0 - it represents the probability of finding the lower-level particle in the ${{\epsilon}_i}$ energy state. 

However, this function is not directly observable or measurable. Measuring this distribution function would take an appliance with a capability to measure $p_i$ vs. ${{\epsilon}_i}$ . The above task may not be performed for a higher-level particle since the particles $m_i$ are indistinguishable. 

The related question is which functions would be observable. To clarify the issue, we rewrite the set of Eqs. ~(\ref{eq_7}) and ~(\ref{eq_8}), transforming the simplest linear combinations of the equations to

\begin{eqnarray}
ln p_i  \pm ln m_i = {\alpha} \pm {\beta}  + {\gamma}{{\epsilon}_i} \pm p_i = 
\nonumber \\  
= {\alpha} \pm {\beta}  + {\gamma}{{\epsilon}_i} \pm 
e^{{\alpha} + {\gamma}{{\epsilon}_i}} 
 \label{eq_11}
\end{eqnarray}

where Eq. ~(\ref{eq_10}) has been utilized. Consequently, the two functions $p_i m_i$ and ${{p_i} \over {m_i}}$ which solve Eqs. ~(\ref{eq_7}) and ~(\ref{eq_8}) must be considered. 

Constants in Eqs. ~(\ref{eq_7}) and ~(\ref{eq_8})  must be defined for further analysis. Consider $\beta$ in Problem 1. The quantity $m_i$ varies from a positive value $e^{{\beta}+e^{\alpha}}$ for ${\epsilon}_i =0$ to $e^{\beta}$ at ${\epsilon}_i \rightarrow \infty$. For large ${\epsilon}_i$, with $\gamma$ being obviously negative, 

\begin{equation}
e^{e^{{\alpha} + {\gamma}{{\epsilon}_i}}} \simeq 1 + e^{{\alpha} + {\gamma}{{\epsilon}_i}} 
 \nonumber
\end{equation}

From the physics standpoint, a large number of high-level particles for a high energy of a lower-level particle contained therein should be an unlikely arrangement, meaning that, as the first step towards treating Problem 1, $\beta = 0$ can be assumed for  ${\epsilon}_i \rightarrow \infty$, and that $m_i$ must tend to 1 for ${\epsilon}_i \rightarrow \infty$, i.e., as stated above,  $\beta = 0$. 

Obviously, the function $p_i m_i$ is potentially observable as it describes the particle number distribution ~(\ref{eq_1}). This function differs substantially from $p_i$ (see Fig. 1 and the corresponding caption where the depicted functions 1-7 are explained in the Caption): 

\begin{figure}
\includegraphics[width=1.0\linewidth]{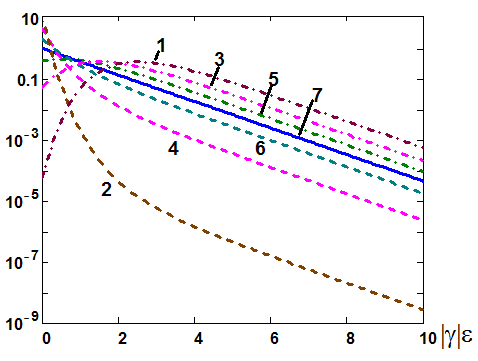}
\caption{\label{Fig.1} Normalized distribution density functions $p_i m_i$ (dashed lines 2, 4, and 6) and ${{p_i} \over {m_i}}$ (dots-and-dashes 1, 3, 5) for various values of the  parameter $\alpha$. The values of the parameter are $\alpha$=2.5 for curves 1 and 2, $\alpha$=1.5 for curves 3 and 4, and $\alpha$=0.5 for curves 5 and 6. The curve $p_i=|\gamma| e^{ \gamma {{\epsilon}_i}}$ is also shown (continuous line 7). It is assumed that $|\gamma| = 1$.  The scale is semi-logarithmic.}
\end{figure}

\begin{equation}
p_i m_i = e^{{\alpha} + {\gamma}{{\epsilon}_i}} e^{e^{{\alpha} + {\gamma}{{\epsilon}_i}}} 
 \label{eq_12}
\end{equation}

The function describing this distribution, normalized to unity, describes the probability density 

\begin{equation}
{(p_i m_i)}_{norm} = |\gamma| {{e^{{\alpha} + {\gamma}{{\epsilon}_i}} e^{e^{{\alpha} + {\gamma}{{\epsilon}_i}}}} \over {e^{e^{\alpha}} -1}} 
 \label{eq_{12a}}
\end{equation}

which is shown as dashed curves in Fig. 1 for $\alpha$ = 0.5, 1.5 and 2.5. As the value of $\alpha / |\gamma|$ increases, e.g. for a decreasing temperature while $\alpha$ stays constant, the value of the function given by Eq.~(\ref{eq_{12a}})  for its argument closed to zero increases sharply against what is seen for line 7, while the asymptote for moderate and large values of ${\epsilon}_i$ drop off with the "same rate" (see Fig. 1). Thus, it appears that the probability density $p_i m_i$ (or the normalized probability density ${(p_i m_i)}_{norm}$ describes condensation of lower-level particles at low energies, and the effect is manifest at low temperatures. The observability of this function seems to materialize best in multi-particle effects such as phase transitions, when "all" of the (higher-level) particles are drawn into the process. Here the situation is not dramatically sensitive to the value of $\alpha$. 

The second function which is a solution to Eq.~(\ref{eq_11}), namely ${{p_i} \over {m_i}}$, is the distribution $p_i$ per one higher-level particle from the $m_i$ set. It is also depicted in Fig. 1. The conditions for its observability are clearly of a different kind than those for $p_i m_i$. It looks that single-particle measurements of the energy spectra (spectroscopic measurements, in a broad sense) are most relevant to the case in terms of the physical meaning. 

Consider the distribution of the lower-level particles over energies "on the average" per single higher-level particle ${{p_i} \over {m_i}}$:

\begin{equation}
{p_i \over m_i} = e^{{\alpha} + {\gamma}{{\epsilon}_i}} e^{- e^{{\alpha} + {\gamma}{{\epsilon}_i}}}  \label{eq_13}
\end{equation}

Obviously, here  $|\gamma|$ plays the role of an "inverse temperature" of the system. Normalization of the distribution given by Eq. ~(\ref{eq_13}) to unity translates into factor of $ |\gamma|/(1-e^{-e^{\alpha}})$ in the equation, and the quantity practically coincides with $|\gamma|$ for moderate or, even more so, large $\alpha$. The function defined by Eq. ~(\ref{eq_13}) peaks for ${\epsilon}_i = {\alpha}/|\gamma|$ and, since $\alpha$ is a Lagrange multiplier for the total number of particles given by Eq. ~(\ref{eq_1}), the physical meaning of $\alpha$ (or, rather, $ {\alpha}/|\gamma|$) may be that it is related to the chemical potential of the lower-level particles $\mu$, with $\alpha =|\gamma| \mu$. It should be noted that the probability density given by Eq. ~(\ref{eq_13}) coincides with the probability density for the Gumbel distribution (type I) ~\cite{gumbel1958statistics}. 

The distribution ${p_i \over m_i}$ given by Eq.  ~(\ref{eq_13}), when normalized to unity, is presented below for convenience of calculations 

\begin{equation}
{({p_i \over m_i})}_{norm} = { |\gamma| \over {1-e^{-e^{\alpha}}}}{e^{{\alpha} + {\gamma}{{\epsilon}_i}} \over e^{e^{{\alpha} + {\gamma}{{\epsilon}_i}}}}
 \label{eq_{13a}}
\end{equation}

Fig. 2 shows the function defined by the above equation for several values of $\alpha$ with different effective temperatures $1 \over {|\gamma|}$.

\begin{figure}
\includegraphics[width=1.0\linewidth]{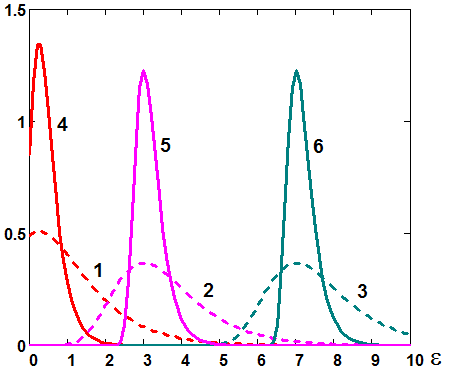}
\caption{\label{Fig.2} Probability distribution density functions ${({p_i \over m_i})}_{norm}$, normalized to unity, vs. ${\epsilon}_i$  ~(\ref{eq_{13a}}) for $\alpha$ = 0.5, 3, and 7. Here $|\gamma|$= 0.5, for the dashed curves 1-3, and $|\gamma|$= 5 for the continuous curves 4-6. The scale is uniform. }
\end{figure} 

As $\alpha$ increases (curves 1 - 3 respectively), the normalizing factor in Eq. ~(\ref{eq_{13a}}) gradually tends to $|\gamma|$. The convergence is practically immediate for low effective temperatures (large $|\gamma|$) and the distribution ${({p_i \over m_i})}_{norm}$ simply translates along the horizontal axis, preserving its shape. It can be seen in Fig. 2 that the maxima become narrow as $|\gamma|$ increases, i.e. as the effective temperature goes down. A sharp maximum of the distribution corresponds to the value of the chemical potential ${\epsilon}_i = \mu = {\alpha}/|\gamma|$. Apparently, the pattern implies that lower-level particles may bond with a higher-level one (or get detached from it) only to occupy energy states close $\mu$. The width of the distribution function decreases if the temperature goes down, and for low temperatures the distribution may become so narrow that no more than a single lower-level particle would be accommodated in such state. 

Therefore, under combinatorial consideration, a two-level hierarchic system of particles, where the lower-level particles are delimited in the higher-level ones, exhibits a split into two sorts of lower-level particles. Tentatively, the two sorts should be observed in different types of experiments specifically examining either collective or individual behavior. Particles of the first sort may condensate into low-energy states. Particles of the second sort should occupy states characterized by the binding energy (or the dissociation energy). In a naively formulated approximation, the states are characterized by the chemical potential of the lower-level particles within the higher-level ones, and, due to the narrowness of the function of probability density of the energy distributions, these happen to be the only states open to the particles.

\section{Problem 1 in terms of distributions of maximal and minimal values. }

Problem 1 can be formulated in other terms such that the solution process and the results would, first, coincide with those demonstrated above, and, secondly, serve to better clarify the entire outcome.  

Assume that Problem 0 is generalized so that there exists a probability density of a distribution over energies $p_i$ for lower-level particles within every higher-level particle (or a distribution inside a box, or a distribution of strings of mushrooms over lengths) from a set $m_i$. Treating the set $m_i$ as a [statistical] {\it sample}, we seek a distribution function (and then - the probability density) of the maximal value $p_i$ over all samples $m_i$. For the purpose, we will follow the procedure outlined in ~\cite{gumbel1958statistics}.

Denote the variable of the function of probability density of the distribution $p_i$ as ${|\gamma|}{{\epsilon}_i}=x$, with the quantity $x$ varying from $0$ to $\infty$ \footnote{Note that the shift of the energy "ground state" of the probability density $p_i$ from zero to a negative value $- {{\epsilon}_0}$ simply leads to a shift of the variable ${{\epsilon}_i}$, which would then vary from $- {{\epsilon}_0}$ to $+ \infty$.  The probability densities obtained for Problem 1 may now be rewritten as ${|\gamma| {(\pm e^{e^{\pm \alpha}} \mp 1)}^{-1}} {e^{\alpha + {\gamma}({{\epsilon}_i} + {{\epsilon}_0})}} {e^{\pm {e^{\alpha + {\gamma}({{\epsilon}_i} + {{\epsilon}_0})}}}}$ for ${(p_i m_i )}_{norm}$ (the upper sign) and $({{p_i} \over {m_i}})_{norm}$ (the lower sign) respectively. The quantities $\alpha$ can have negative values which, compared to those considered in the previous Section, should be shifted by $-{{\epsilon}_0}$ to obtain the same probability densities. All the results laid out above would remain unchanged.}.  The distribution function $F(x)$ is, naturally, $F(x)=1-e^{(-x)}$. The probability of encountering a maximal value in a sample $m_i$ is determined by the expression ~\cite{gumbel1958statistics}
  
\begin{equation}
m_i [ 1 - F ({x_{im}})] = 1
 \label{eq_{14}}
\end{equation}

where $x_{im}$ is the characteristic maximal value. It follows from Eq. ~(\ref{eq_{14}}) that 

\begin{equation}
x_{im} = {\ln}m_i  \; and/or \;  {{e^{x_{im}}} \over {m_i}} = 1
 \label{eq_{15}}
\end{equation}

The distribution function ${F_m} (x)$ of the greatest value over the entire $m_i$ sample is, obviously, the function $ F(x)$ to the power $m_i$ [6]. Using the second of Eqs. ~(\ref{eq_{15}}) and a transition to a limit, we arrive at 

\begin{equation}
{F_m}(x) = (1 - {{e^{-(x - x_{im})}} \over {m_i}})^{m_i} \rightarrow {e^{-e^{-(x - x_{im})}}}
 \label{eq_{16}}
\end{equation}
while $m_i \rightarrow \infty$. Differentiating the limiting function of the distribution as defined by Eq. ~(\ref{eq_{16}}), one finds the probability density ${p_i} \over {m_i}$, where $x_{im} = \alpha$, which corresponds to the normalization of the Boltzmann-Gibbs distribution

\begin{equation}
{m_i} e^{- x} = e^{-(x - {\ln} m_i)}
 \nonumber
\end{equation}
by the number of higher-level particles $m_i$. 

Next, consider the function $p_i m_i$ in line with an analogous scheme. In contrast to $1 \over {m_i}$, the function $m_i$ is not a distribution function of any random variable. At the same time, it is known ~\cite{gumbel1958statistics} that the function of the probability density of the distribution of minimal values $p_i$ in the sample $m_i$ (or the distribution over energies of the lower-level particles within the higher-level ones) is ${m_i} e^{-m_i x}$  (we retain the notation ${|\gamma|}{{\epsilon}_i}=x$). 

Let us consider the case of zero values of $\alpha$. We develop a new function of the probability density according to the following scheme: initially we adopt the above distribution of the minimal value, as it can occur in any specific case over the entire sample $m_i$ (within any of the higher-level particles). Since it is irrelevant in which realization (within which higher-level particle) this would happen, we ascribe to this probability density the statistical weight of all combinatorial permutations with repetitions, i.e.  $1 \over {(m_i !)}$. The term $ {[{(m_i - 1)} !]}^{-1} e^{-m_i x} = e^{-x} {[{(m_i - 1)} !]}^{-1} e^{- (m_i -1 ) x}$ is to serve as the first term of the sum to be compiled, which will represent the final value of a function of probability density.

The probability density of the minimal value in the remaining $m_i - 1$ realizations (the remaining higher-level particles) makes  $({m_i} - 1) e^{- (m_i -1) x}$. Similarly, the weight to be ascribed to this function is $1 \over {(m_i - 1) !}$, the result being another term of the sum and a contribution to the probability density function $ {[{(m_i - 2)} !]}^{-1} e^{-(m_i -1) x} = e^{-x} {[{(m_i - 2)} !]}^{-1} e^{- (m_i -2 ) x}$. The procedure should be repeated a total of  $m_i -1$ times.  With a limiting transition 

\begin{equation}
e^{-x} {[{(m_i - k')} !]}^{-1} e^{- (m_i -k' ) x} \rightarrow e^{-x} {( k ! )}^{-1} e^{- k x}
 \label{eq_{??}}
\end{equation}
the probability density functions add up to the required probability density function (see Fig. 3):
\begin{figure}
\includegraphics[width=1.0\linewidth]{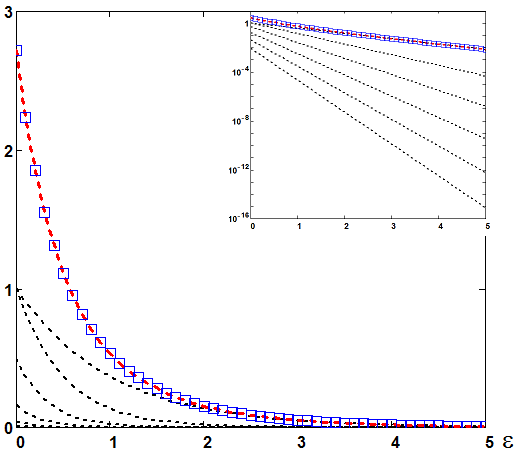}
\caption{\label{Fig.3} The function $e^{-x} e^{e^{-x}}$ (the curve consisting of open squares) compared to the sum of the functions $e^{-x} {\sum}_{k=0}^5 {{e^{-kx}} \over {k !}}$ (the upper bold dashed curve). The six narrow dashed curves below represent the functions $e^{-x} {{e^{-kx}} \over {k !}}$, with $ k$ ranging from 0 to 5. The main plot, uniform coordinates. The inset - same in semilogarithmic scale.}
\end{figure}

It is possible to do since we take different group of particles each time. The final probability distribution density function is now the sum of partial ones:
\begin{equation}
f ( x) = {\lim}_{m_i \rightarrow \infty}  e^{-x} {\sum}_{k=0}^{m_i -1} {{e^{-kx} \over {k !}} \rightarrow e^{-x} e^{e^{-x}}
 \label{eq_{17}}}
\end{equation}

The function $p_i m_i$ exactly equals $f(x)$ for $\alpha = 0$. Practically, $f(x)$ represents the probability density of minimal values of $p_i$, obtained in every realization of the $m_i$ sample (the energy of the lower-level particle has a minimal value within a higher-level particle provided that the values are also minimal in other realizations). Apparently, this is direct evidence of the fact that $f(x)$ (or $p_i m_i$) describes a kind of collective behavior of the higher-level particles as well as of the lower-level particles within them.  

There is significant difference between $f(x)$ and $p_i \over m_i$. The latter describes the density of the probability distribution of a "normal" particle having a certain energy, which happens to be maximal over a set of higher-level particles. The result, namely the maximal value distribution, is the same in every higher-level particle (in every realization of the sample), though it is clearly determined by the sample as a whole. For $f(x)$ (or, generally, for the function $p_i m_i$) the result for every realization is fundamentally different. This may mean that the system of higher-level particles with the given lower-level particles comprises a certain collective system, where the lower-level particles within a higher-level particle are somehow affected by each others presence. Thus, delimitation in this case additionally defines a certain (potentially collective) interaction between species. The above is a stronger statement than a one describing "normal" particles with the function$p_i \over m_i$ would take, as here the delimitation only leads to the observation of maximal values of the $p_i$ distribution.  

A calculation of the probability density of lower-level particles over energy for a sample of the higher-level particles produces exactly the same result as the combinatorial calculation. From the standpoint of statistical mechanics, only states with maximal energy or conditionally (see above) minimal energy of the lower-level particles may emerge as the most probable (observable). An effective Pauli principle arises in the former case. 
 
\section{Nature of the effective Pauli principle. }

Several facts related to the above calculations are indicative for an explanation of the effective Pauli principle nature. First, the function $p_i \over m_i$ in the framework for maximal value describes the probability density  of the maximal value $p_i$ over the sample $m_i$ or, which is the same in terms of particles, the probability density for registering a lower-level particle in a maximal energy state over the set of higher-level particles.  This perspective affords only one lower-energy particle in a definite energy state, which is a manifestation of the effective Pauli principle. 

Secondly, when the temperature goes down both types of calculations establishing the probability density $p_i \over m_i$ point to a narrowing of the distributions (Fig. 2). For low temperatures, the distribution can become so narrow that now more than one lower-level particle would "fit" into it. A legitimate question in this connection is what states would be open to the remaining lower-level particles.

Fig. 4 illustrates a model situation where an ensemble of higher-level particles with various $\alpha$ (various chemical potentials for corresponding groups of the lower-level particles) is available. 

\begin{figure}
\includegraphics[width=1.0\linewidth]{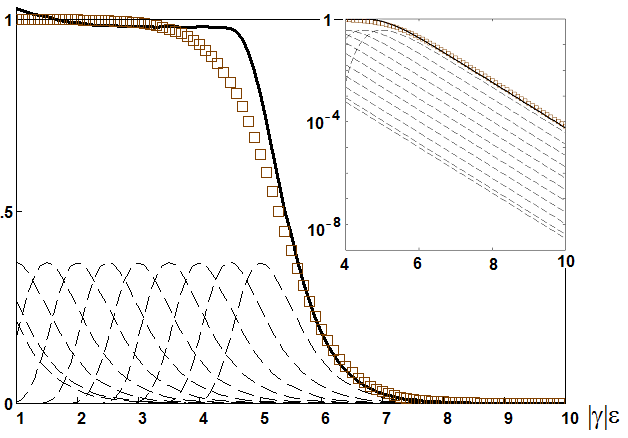}
\caption{\label{Fig.4}The functions $p_i \over m_i$  (dashed curves) with the values of $\alpha$ from 0.5 to 5 (over 0.5), and a continuous curve, representing the sum of these functions. The open squares show the Fermi functions $1 \over {exp{(|\gamma| \epsilon - \alpha)}+1}$ for $\alpha \simeq 5.3$. The inset shows in semilogarithmic scale the values of the functions for large values of their arguments, explicating the exponential asymptotic of the summary function. $|\gamma| = 1$.   }
\end{figure}

In the case, the remaining lower-level particles are accommodated in the energy states of the corresponding distributions (dashed curves in Fig. 4) with identical maxima. If they are distributed densely, with ${\alpha}_{min} < \alpha < {\alpha}_{max}$  (${\alpha}_{min}$ may equal zero), the summary probability density (dashed curve in Fig. 4) over energy appears to be constant wherever $\epsilon < {{\alpha}_{max} \over {|\gamma|}}$, and then decreases exponentially (see the inset in Fig. 4). Such summary distribution would agree well with the Fermi distribution for $ \mu \simeq {\mu}_{max} = {{\alpha}_{max} \over {|\gamma|}}$. Consequently, the Fermi distribution density for all delimited particles (quantitatively, the delimitation is defined by some energy - the Fermi energy which is close to the chemical potential for the isolated set of particles) can be represented by a sum of the probability densities of the distributions of groups of particles, every one of which has its unique value of the chemical potential. The above may also provide a key to the interpretation of the origin of the effective Pauli principle. 

For the same purpose, it is enough to note, without citing the complexities which Fig. 4 reflects, that some energy of a lower-level particle would always be registered in single-particle measurements of the state of the system of species, and that for low temperatures the energy may be unique, as other values do not fit under curves 4-6 in Fig. 2 due to the spacing between levels of the discrete energy spectrum. So, one energy state would correspond to only one particle, which may be interpreted as a case of the effective Pauli principle. 

At this point, the principle can be formulated as follows. Lower-level particles delimited in higher-level ones have, as the most probable, a probability density of a distribution over energies shaped as a narrow bell-shaped curve. On top of the fact that only one particle "fits" into the distribution at low temperature, the function is also a distribution of lower-level particles having a maximal energy within a higher-level particle.The probability fo find two particles in the state with maximal energy is clearly extremely small. Thus, only one delimited particle can be found in one (highest) energy state - and namely these particles are observed.  

\section{Problem 2}

(see ~\cite{romanovsky2021statistical}). As it is noted in the Introduction, a formulation of the problem can address the distribution over energies of lower-level particles, contained in a set of intermediate-level particles, which, in their turn, are enclosed in higher-level particles. The particles are indistinguishable on all the three levels \footnote{In terms of the generalized "mushroom drying" problem, this would mean a calculation of the lengths of mushroom strings which are organized in bunches of different volumes. Not only the strings, but also the bunches - not to mentions the mushrooms - are indistinguishable.}. Denote the number of lower-level particles contained in an intermediate-level particle as $p_i$, the number of intermediate-level particles with this number of lower-level particles as $m_i$, and the number of higher-level particles in which intermediate-level particles in the amount of $m_i$ are packed as $t_i$. The total number of permutations of lower-level particles over all intermediate-level particles, and of those - over the higher-level particles is

\begin{equation}
W_{21} = {N! \over{{{\prod}_{i=1}^n}{(p_i !)}^{m_i t_i}{{\prod}_{i=1}^n}{{(m_i !)}^{t_i}}{{\prod}_{i=1}^n}{(t_i !)}}} 
 \label{eq_18}
\end{equation} 

Here again $n$ is the full number of distinct arrangements of the lower-level particles over the intermediate-level particles, and of the latter - over the higher-level particles. Naturally, the total number of lower-level particles is

\begin{equation}
N = {{\sum}_{i=1}^n}{m_i}{p_i}{t_i} 
 \label{eq_19}
\end{equation}

the total number of intermediate-level particles is

\begin{equation}
M = {{\sum}_{i=1}^n}{m_i}{t_i} 
 \label{eq_20}
\end{equation}

and the total number of higher-level particles is

\begin{equation}
T = {{\sum}_{i=1}^n}{t_i} 
 \label{eq_21}
\end{equation}

The total energy is given by

\begin{equation}
E = {{\sum}_{i=1}^n}{m_i}{p_i}{t_i}{{\epsilon}_i} 
 \label{eq_22}
\end{equation}

Following the general rule, consider the logarithm of the number of permutations

\begin{eqnarray}
ln W_{21} = ln N! - {{\sum}_{i=1}^n}{m_i}{t_i}{ln (p_i !)} - 
\nonumber \\  
- {{\sum}_{i=1}^n}{t_i}{ln (m_i !)} - {{\sum}_{i=1}^n}{ln (t_i !)}
 \label{eq_23}
\end{eqnarray}

Accordingly, the quantity given by Eq. ~(\ref{eq_23}) is subject to variation under the constraints given by Eqs. ~(\ref{eq_19}) - ~(\ref{eq_22}) with Lagrange multipliers $\alpha, \beta, \gamma, \Delta$, in analogy with Eq. ~(\ref{eq_6}):

\begin{equation}
ln W_{2} = ln W_{21} + \alpha N + \beta M + \gamma E + \Delta T
 \label{eq_24}
\end{equation}

The variables for variation here are $p_i, m_i, t_i$. Variation over $p_i$ results in an equation identical to Eq. ~(\ref{eq_7}) (leading to a solution same as for Problem 0):

\begin{equation}
ln p_i = {\alpha} + {\gamma}{{\epsilon}_i} 
 \label{eq_25}
\end{equation}

The variation of Eq. ~(\ref{eq_24}) over $m_i$ combined with Eq. ~(\ref{eq_21}) produces an equation identical to Eq. ~(\ref{eq_8}):

\begin{equation}
ln m_i = {\beta} + p_i 
 \label{eq_26}
\end{equation}

Finally, the variation of Eq. ~(\ref{eq_24}) over $t_i$, combined with Eqs. ~(\ref{eq_25}) and ~(\ref{eq_26}), results in a third equation for Problem 2

\begin{equation}
ln t_i = {\Delta} + m_i 
 \label{eq_27}
\end{equation}

Certain remarks are due concerning the quantities $\beta$ and $\Delta$. The latter quantity features in all of the equations within an exponential factor, being potentially necessary only in the case of a special normalization, while it can be set to zero in the lowest approximation, though it would have been natural to assume that  $\Delta = -1$, so that $t_i$ would tend to $1$ for ${{\epsilon}_i} \rightarrow \infty$ (see above).  The quantity $\beta$ in Problem 2 can, for a start, be set equal to $\beta$ from Problem 1, which is also zero. The case of $\beta$ not equal to zero will be treated further on. 

We compile, as in the case of Problem 1, a simple linear combination of Eqs.  ~(\ref{eq_25}) - ~(\ref{eq_27}):

\begin{eqnarray}
ln p_i  \pm ln m_i \pm ln t_i = 
= {\alpha} \pm {\beta}  \pm \Delta +  {\gamma}{{\epsilon}_i} \pm p_i \pm m_i
\nonumber 
\end{eqnarray}

and make use of Eqs. ~(\ref{eq_25}) and ~(\ref{eq_26}). The result is 

\begin{eqnarray}
ln p_i  \pm ln m_i \pm ln t_i = 
\nonumber \\  
= {\alpha} \pm {\beta}  \pm \Delta +  {\gamma}{{\epsilon}_i} \pm 
\nonumber \\  
\pm e^{{\alpha} + {\gamma}{{\epsilon}_i}} \pm  e^{\beta + e^{{\alpha} + {\gamma}{{\epsilon}_i}}}
 \label{eq_28}
\end{eqnarray}

In full similarity to Problem 1, the solutions to Eq. ~(\ref{eq_28}) and the distribution functions in the case considered are $p_i m_i t_i, {{p_i t_i} \over m_i}, {p_i \over {m_i t_i}}$, as well as ${{p_i m_i} \over t_i}$. The first of the functions describes multiparticle measurements with the effect of particle condensation at low energies, which is more intense than in problem 1 (see Fig. 5).  The second function describes the effect of condensation of the lower-level particles, but it is less pronounced than in the case of the first of the functions, as illustrated by the comparison of Curves 3 and 4 against Curves 1 and 2 in Fig. 5. The third one describes a certain function measured per intermediate-level particle and having a sharp maximum. Here, however, the maximum does not correspond to the value of $\alpha$ even for zero $\beta$, but exceeds it approximately by $1$ for ${p_i \over {m_i t_i}}$ under the conditions of Fig. 5 (this is equivalent to approximately one unit of temperature), as demonstrated by Curves 5 and 6. 

\begin{figure}
\includegraphics[width=1.0\linewidth]{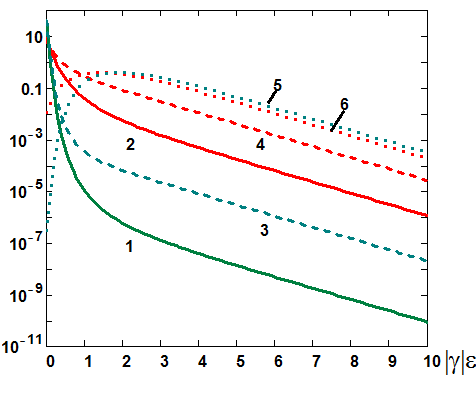}
\caption{\label{Fig.5} Probability distribution density functions $p_i m_i t_i$, normalized to unity, vs ${\epsilon}_i$    (Curve 1 for $\alpha$ = 1 and Curve 2 for $\alpha$ = 0.5), ${{p_i t_i} \over m_i}$ (Curve 3 for $\alpha$ = 1 and Curve 4 for $\alpha$ = 0.5), ${p_i \over {m_i t_i}}$ (Curve 5 for $\alpha$ = 1 and Curve 6 for $\alpha$ = 0.5). Uniform scale, $|\gamma|$ is set to 1. }
\end{figure}

Finally, ${{p_i m_i} \over t_i}$ is the probability density for the distribution ${e^{e^{\beta}}} \over t_i$  , if the energy, which is the argument here, is extrapolated to cover negative values. The normalized distribution of the probability density for the positive values of the argument is 

\begin{equation}
{({{p_i m_i} \over t_i})}_{norm} = { |\gamma| \over {{1 \over e}-e^{-e^{e^{\alpha}}}}}
{{e^{{\alpha} + {\gamma}{{\epsilon}_i}}e^{e^{{\alpha} + {\gamma}{{\epsilon}_i}}}} \over e^{e^{e^{{\alpha} + {\gamma}{{\epsilon}_i}}}}}
 \label{eq_29}
\end{equation}

As in the case of Problem 1, the function defined by Eq. ~(\ref{eq_29}) is a result of measuring an average value per one higher-level particle. 

Fig. 6 illustrates the behavior of the normalized density of the probability distribution given by Eq. ~(\ref{eq_29}) depending on the Lagrange multipliers $\alpha$ and $\beta$.

\begin{figure}
\includegraphics[width=1.0\linewidth]{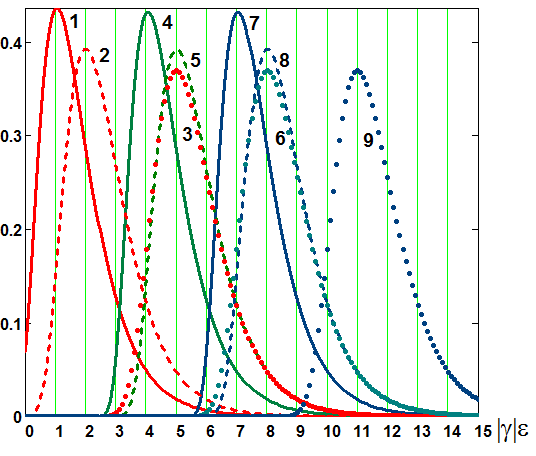}
\caption{\label{Fig.6}  Probability distribution density functions ${({{p_i m_i} \over t_i})}_{norm}$ given by Eq. ~(\ref{eq_29}), normalized to unity, vs ${\epsilon}_i$.   Continuous curves 1, 4 and 7 for $\alpha$ = 0, $\beta$ = 1; $\alpha$ = 3, $\beta$ = 1;  $\alpha$ = 6, $\beta$ = 1 correspondingly. Dashed curves 2, 5 and 8 for $\alpha$ = 0, $\beta$ = 2; $\alpha$ = 3, $\beta$ = 2;  $\alpha$ = 6, $\beta$ = 2 correspondingly, point curves 3, 6 and 9 for $\alpha$ = 0, $\beta$ = 5; $\alpha$ = 3, $\beta$ = 5;  $\alpha$ = 6, $\beta$ = 5 correspondingly. For all curves, $|\gamma| = 1$.   }
\end{figure} 
The maximum of ${{p_i m_i} \over t_i}$ for $\beta > 1 $ coincides with high precision with $\alpha + \beta$, so that the quantity $\beta$ may be "naively" interpreted as the chemical potential on one lower-level particle within the higher-level particle (for $\alpha$, the interpretation follows from Problem 1). It becomes obvious in the light of such reading that the full chemical potential of a lower-level particle in the higher-level particle, in the framework of Problem 2, comprises the two chemical potentials, namely, of the lower-level particle within the intermediate-level one and of the intermediate-level particle in the higher-level one. The fact can be seen clearly in Fig. 6 as the maxima of the probability densities shown by curves 2, 3, 5, 6, 8, and 9 correspond to the sum $\alpha + \beta$.

For $0 \leq \beta \leq 1$, the maximum of ${{p_i m_i} \over t_i}$ only slightly exceeds $\alpha + \beta$, as it can be seen from curves 1, 4, and 7 in Fig. 6. For other curves, the coincidence takes place with the precision up to ${e^{-2(\alpha + \beta)} \over 2}$. It also follows from Fig. 6 that the different chemical potentials (proportional to $\alpha$ and $\beta$) yielding identical sums define probability densities which are very similar (cures 3 and 5, as well as 6 and 8 in Fig. 6).

The calculations of the probability distribution functions of maximal and minimal values that would serve to confirm and generalize the combinatorial calculations are impossible to carry out in the case of Problem 2. Only one of the four functions of probability density, namely ${{p_i m_i} \over t_i}$, can be obtained from the probability distribution given by ${e^{e^{\beta}}} \over t_i$ , but this is not achieved by either finding the maximal value distribution or by constructing the distribution of the collective minimal value.

Thus, Problem 2, which describes a three-level hierarchic system, highlights the splitting of the modes of behavior of the lower-level particles into four types.  Two of them describe the condensation of the particles into states with low energies. One type of particles, described by the probability density ${{p_i m_i} \over t_i}$, is closed to the type of particles described by the function ${{p_i} \over m_i}$ in Problem 1. In a naive representation, this type (like the similar one in Problem 1), possibly corresponds to "normal" bound particles.

It is unclear what the fourth probability density ${{p_i} \over {m_i t_i}}$ describes.

\section{On the negligible probability to find negative-energy particles}

Let us note that, if the argument $x$ is allowed to vary over the range from $- \infty$ to $+ \infty$, the function ${1 \over m_i} = e^{-e^{-(x-x_{im})}}$ exactly becomes a probability distribution of some random quantity. Let us keep the relation $x =|\gamma| {\epsilon}_i$. It is interesting that for a definite, not too large value $x_im \simeq \alpha = |\gamma| \mu \geq 6$, the lower-level particles with negative energies are totally absent (the probability of finding at least one particle makes a vanishing quantity $ \sim 10^{-180}$, as illustrated by Fig. 7. 

\begin{figure}
\includegraphics[width=1.0\linewidth]{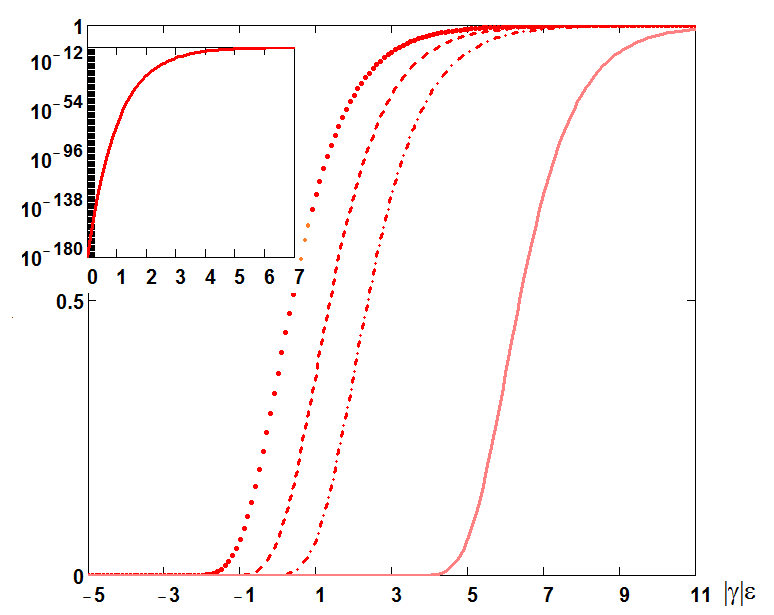}
\caption{\label{Fig.7} The distribution function $1 \over m_i$ vs energy ${\epsilon}_i$ for $\alpha = |\gamma| \mu $ equal to 0 for the point curve, 1 for the dashed curve, 2 for the dot-dashed curve, and 6 (more exactly, $6.025$) for solid curve. For $\alpha = |\gamma| \mu = 6.025$, the function for zero value of its argument would make  $~ 10^{-180}$. The event of the appearance of a particle with zero energy with the probability of $~ 1$ would have taken around $10^{180}$ higher-level particles, which is likely more than is available across the universe. Uniform coordinates, $|\gamma|$ is set to $1$. The inset - the solid curve with $\alpha = 6.025$, the semilogarithmic scale   }
\end{figure} 

Since the delimitation of lower-level particles within lower-level ones is achieved due to an interaction of some nature, it implies non-zero values of the chemical potential. Apparently, the interaction is responsible for the absence of particles in negative energy states. 

The following remark is due in connection with the functions arising in Problem 2. The function  

\begin{equation}
{ e^{e^{\beta}} \over t_i} = e^{e^{\beta}} e^{-e^{{\beta} +{e^{\alpha - x}}}}
 \label{eq_30}
\end{equation}

is a distribution of some random variable provided that the argument $x = |\gamma| {\epsilon}_i$  varies from $- \infty$ to $+ \infty$ (Fig. 8). For $\alpha =0$, this function, depicted as Curve 3, has the value for zero argument $< 10^{-180}$ already for $\beta \geq 3.4$. If the span of the argument starts with zero (or some other value corresponding to the ground state of a lower-level particle), differentiation easily produces the function ${p_i m_i} \over t_i$ (Fig. 6).

\begin{figure}
\includegraphics[width=1.0\linewidth]{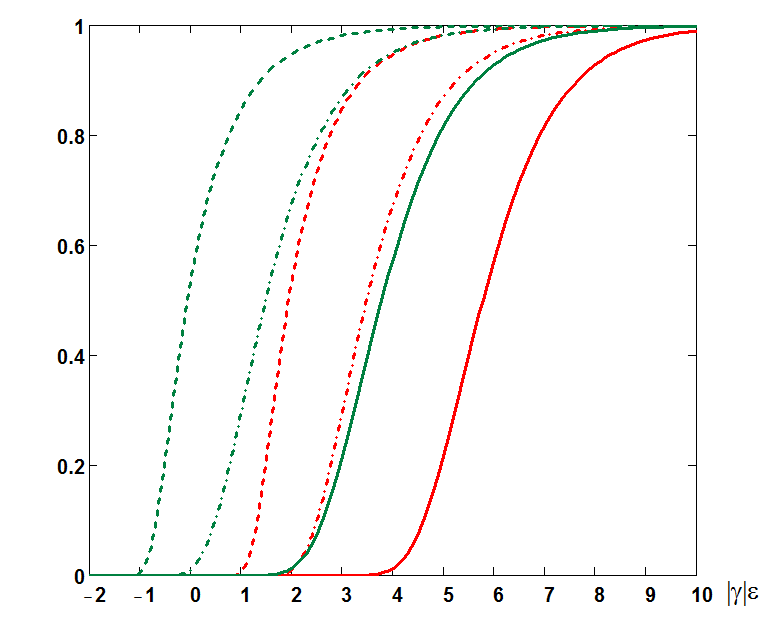}
\caption{\label{Fig.8} Probability distributions given by Eq. ~(\ref{eq_29}) for various values of $\alpha$ and $\beta$: curve 1, $\alpha$=0 $\beta$=-1, curve 2, $\alpha$=0 $\beta$=1, curve 3 $\alpha$=0 $\beta$=3.4, curve 4 $\alpha$=2 $\beta$=-1, curve 5 $\alpha$=2 $\beta$=1, curve 6 $\alpha$=2 $\beta$=3.4. Uniform coordinates, $|\gamma|$ is set to 1.   }
\end{figure} 

The function of probability density also describes lower-level particles, which cannot exist in negative-energy states. At least one of the chemical potentials not being zero is sufficient to warrant the statement (see above). 

\section{Discussion of results.}

The results for Problems 1 and 2 can be interpreted as depended on the method of observation. All observations in spectroscopic experiments (not necessarily limited to optics, but generally dealing with the measurements of the energy spectra) produce registrations of distributions pertinent to a higher-level particle. The main result for Problem 1 is the finding of the maximum of the probability density function for a distribution of particles over energies for a one-particle observation. In the framework of the simplest descriptive approach, the maximum coincides with the chemical potential of lower-level particles within the higher-level particles, i.e. the energy of binding/release of a lower-level particle to/from likewise species within a higher-level particle. The simple treatment considers all particles with energies below the chemical potential as bound. 

As it has been already stated, for low effective temperatures the density of the probability to find a lower-level particle in the state ${\epsilon}_i$ via a one-particle measurement exhibits an extremely narrow peak. This translates into no more than one particle occupying the state at low (and yet lower) temperature. The result can be understood as an effective Pauli principle with one energy state taken by one particle of a certain sort, corresponding to the minus sign in the left-hand side of Eq. ~(\ref{eq_11}).  A modification of Problem 1 giving rise to a set of functions ${p_i \over m_i}$ with various $\alpha$ yields, as the functions add up, a function close to the Fermi distribution. This is yet another potential realization of the effective Pauli principle which emerges from the solution to Problem 1. 

It is furthermore established that the function of the probability density ${p_i \over m_i}$ describes the distribution of the maximal value of $p_i$ in $m_i$ sample, or, which is the same in terms of particles, the probability density of finding a lower-level particle in a maximal energy state over the entire higher-level sample. Since, obviously, only one lower-level particle can occupy a maximal energy state, the situation amounts to a direct realization of the Pauli principle. 

The behavior of another observable function in Problem 1, namely $p_i m_i$, is also governed by the effective temperature. Already for moderate values $\alpha \simeq 6$ the share of particles subject to the normalized probability density distribution given by Eq. Eq. ~(\ref{eq_{12a}}) with energies beyond the corresponding chemical potential becomes $\simeq 10^{-180}$, i.e. such (free) particles are unlikely to exist in the universe. The formulation of the problem about the probability density function for finding lower-level particles with minimal energies brings about the same result. A nave interpretation of the results of treating Problem 1 with two distribution functions $p_i m_i$ and ${p_i \over m_i}$ is that we deal with particles of two sorts - pseudobosons in the case of the first function (as it describes condensation in low-energy states for low temperature) and pseudofermions (as, for low temperatures, one energy state, seen to correspond to the maximal energy, is occupied by only one particle). 

The same interpretation of the outcome for Problem 2 reveals 4 sorts of particles. Pseudobosons with an even more pronounced condensation effect are described by the function $p_i m_i t_i$, the particles with energies beyond the chemical potential being totally absent. Similar results are sustained by the function ${p_i t_i} \over m_i$, but it reflects a weaker condensation effect. In contrast, the higher-level particles described by the function $ p_i \over {m_i t_i}$ contain lower-level particles in energy states beyond the chemical potential, while the function  ${p_i m_i} \over t_i$ appears to describe some "normal" particles as the total chemical potential of a lower-level particle within an intermediate-level one, and of the latter - within a higher-level particle combine the chemical potentials of the lower-level particle in an intermediate-level particle and of the intermediate-level particle in a higher-level particle.  

The most nave interpretation of the extremely low values of the of the probability distributions $1 \over m_i$  and ${e^{e^{\beta}}} \over t_i$  for negative values of the normalized energy (in the energy range from $- \infty$ to $+ \infty$), may be that the lower-level particles with negative energies (or with energies below a certain negative threshold ${{\epsilon}_0}$ - see footnote 3) may not combine into higher-level particles, or that particles with energies below the ground state cannot exist.
{\footnote{It pulls to treat this fact to absent of antiparticles}}. 
The reason behind this, i.e. the reason for such low values of the distribution function for negative arguments, is the delimitation of some of the particles within others, or an implicit assumption of an interaction taking place.  

\section{Acknowledgments.} The author acknowledges useful discussions of results with I.V. Kolokolov and O.B. Shiryaev.


\end{document}